\documentclass[journal=ACSNano,manuscript=article]{achemso}
\setkeys{acs}{keywords = true}
\setkeys{acs}{etalmode=truncate}
\setkeys{acs}{maxauthors=20}
\usepackage{chemformula} % Formula subscripts using \ch{}
\usepackage[T1]{fontenc} % Use modern font encodings

\usepackage{hyperref}
\hypersetup{%
	citecolor=[RGB]{0,0,0},%
	colorlinks=true,%
	linkcolor=[RGB]{0,0,0}%
}

\usepackage{lineno}

\DeclareCaptionLabelFormat{myformat}{#2}
\author{Changhua Bao}
\affiliation{Department of Physics, Tsinghua University, Beijing 100084, People's Republic of China}
\alsoaffiliation{State Key Laboratory of Low-Dimensional Quantum Physics, Tsinghua University, Beijing 100084, People's Republic of China}

\author{Fei Wang}
\affiliation{Department of Physics, Tsinghua University, Beijing 100084, People's Republic of China}
\alsoaffiliation{State Key Laboratory of Low-Dimensional Quantum Physics, Tsinghua University, Beijing 100084, People's Republic of China}

\author{Haoyuan Zhong}
\affiliation{Department of Physics, Tsinghua University, Beijing 100084, People's Republic of China}
\alsoaffiliation{State Key Laboratory of Low-Dimensional Quantum Physics, Tsinghua University, Beijing 100084, People's Republic of China}

\author{Shaohua Zhou}
\affiliation{Department of Physics, Tsinghua University, Beijing 100084, People's Republic of China}
\alsoaffiliation{State Key Laboratory of Low-Dimensional Quantum Physics, Tsinghua University, Beijing 100084, People's Republic of China}

\author{Tianyun Lin}
\affiliation{Department of Physics, Tsinghua University, Beijing 100084, People's Republic of China}
\alsoaffiliation{State Key Laboratory of Low-Dimensional Quantum Physics, Tsinghua University, Beijing 100084, People's Republic of China}

\author{Hongyun Zhang}
\affiliation{Department of Physics, Tsinghua University, Beijing 100084, People's Republic of China}
\alsoaffiliation{State Key Laboratory of Low-Dimensional Quantum Physics, Tsinghua University, Beijing 100084, People's Republic of China}

\author{Xuanxi Cai}
\affiliation{Department of Physics, Tsinghua University, Beijing 100084, People's Republic of China}
\alsoaffiliation{State Key Laboratory of Low-Dimensional Quantum Physics, Tsinghua University, Beijing 100084, People's Republic of China}

\author{Wenhui Duan}
\affiliation{Department of Physics, Tsinghua University, Beijing 100084, People's Republic of China}
\alsoaffiliation{State Key Laboratory of Low-Dimensional Quantum Physics, Tsinghua University, Beijing 100084, People's Republic of China}
\alsoaffiliation{Frontier Science Center for Quantum Information, Beijing 100084, People's Republic of China}
\alsoaffiliation{Institute for Advanced Study, Tsinghua University, Beijing 100084, People’s Republic of China}

\author{Shuyun Zhou}
\affiliation{Department of Physics, Tsinghua University, Beijing 100084, People's Republic of China}
\alsoaffiliation{State Key Laboratory of Low-Dimensional Quantum Physics, Tsinghua University, Beijing 100084, People's Republic of China}
\alsoaffiliation{Frontier Science Center for Quantum Information, Beijing 100084, People's Republic of China}
\alsoaffiliation{E-mail: syzhou@mail.tsinghua.edu.cn}

\title[An \textsf{achemso} demo]
  {Light-induced ultrafast glide-mirror symmetry breaking in black phosphorus}

\keywords{Floquet engineering, light-induced glide-mirror symmetry breaking, black phosphorus, TrARPES, fully gapped nodal ring}

\begin{document}
%\linenumbers
%%%%%%%%%%%%%%%%%%%%%%%%%%%%%%%%%%%%%%%%%%%%%%%%%%%%%%%%%%%%%%%%%%%%%
%% The "tocentry" environment can be used to create an entry for the
%% graphical table of contents. It is given here as some journals
%% require that it is printed as part of the abstract page. It will
%% be automatically moved as appropriate.
%%%%%%%%%%%%%%%%%%%%%%%%%%%%%%%%%%%%%%%%%%%%%%%%%%%%%%%%%%%%%%%%%%%%%

%%%%%%%%%%%%%%%%%%%%%%%%%%%%%%%%%%%%%%%%%%%%%%%%%%%%%%%%%%%%%%%%%%%%%
%% The abstract environment will automatically gobble the contents
%% if an abstract is not used by the target journal.
%%%%%%%%%%%%%%%%%%%%%%%%%%%%%%%%%%%%%%%%%%%%%%%%%%%%%%%%%%%%%%%%%%%%%
\begin{abstract}
Symmetry breaking plays an important role in fields of physics, ranging from particle physics to condensed matter physics. In solid-state materials, phase transitions are deeply linked to the underlying symmetry breakings, resulting in a rich variety of emergent phases.  Such symmetry breakings are often induced by controlling the chemical composition and temperature or applying an electric field and strain, etc. In this work, we demonstrate an ultrafast glide-mirror symmetry breaking in black phosphorus through Floquet engineering. Upon near-resonance pumping, a light-induced full gap opening is observed at the glide-mirror symmetry protected nodal ring, suggesting light-induced breaking of the glide-mirror symmetry.  Moreover, the full gap is observed only in the presence of the light-field and disappears almost instantaneously ($\ll$100 fs) when the light-field is turned off, suggesting the ultrafast manipulation of the symmetry and its Floquet engineering origin. This work not only demonstrates light-matter interaction as an effective way to realize ultrafast symmetry breaking in solid-state materials, but also moves forward towards the long-sought Floquet topological phases.
\end{abstract}

%%%%%%%%%%%%%%%%%%%%%%%%%%%%%%%%%%%%%%%%%%%%%%%%%%%%%%%%%%%%%%%%%%%%%
%% Start the main part of the manuscript here.
%%%%%%%%%%%%%%%%%%%%%%%%%%%%%%%%%%%%%%%%%%%%%%%%%%%%%%%%%%%%%%%%%%%%%
~
\renewcommand{\thefigure}{Figure~\arabic{figure}}
\setcounter{figure}{0}

Symmetry breaking describes the transition from a symmetric phase into a less symmetric but ordered phase\cite{gross1996role,basov2017towards,SunNRP2021} in the Landau's paradigm\cite{Landau1937},  such as charge density wave induced by translational symmetry breaking\cite{grunerCDWRev1988}. Beyond Landau's paradigm, the topological properties of solid-state materials are still tightly linked to the symmetry\cite{RevModPhys.82.3045,RevModPhys.83.1057}, where the featured band crossing nodes are protected by different kinds of symmetries, such as time-reversal and inversion symmetry protected Dirac nodes\cite{PhysRevB.83.205101,NB2014Sci, CA2014NM}. In addition to the symmorphic symmetries including point group symmetry and translational symmetry, the nonsymmorphic symmetry, which involves a point group symmetry operation followed by a fractional lattice translation\cite{bradley2009mathematical}, further enriches the topological phases. Examples of topological phases with nonsymmorphic symmetry include topological nodal lines or nodal rings\cite{Fu2015PRB,KanePRL2015,Nodalchain2016,NodallineRev2018}, which are robust against the hybridizations\cite{Fu2015PRB,KanePRL2015} and could host various types of novel quasiparticles including drumhead fermions\cite{Wen2015PRB}, M\"obius fermions\cite{Mobius2015PRB} and hourglass fermions\cite{BernevigNat2016}.

Symmetry breaking can be realized by external fields such as static electric fields\cite{basov2017towards,SunNRP2021}, and even oscillating light-field, which allows to break the symmetry on an ultrafast timescale\cite{okaRev2019,Lindner2020NRP,Chris2021rev,Sentef2021RMP,ZhouNRP2021,qi2022traversing,ahn2023ultrafast}. For instance,  breaking inversion symmetry with light-field to tailor superconductivity\cite{yang2019lightwave} or topological states\cite{sie2019ultrafast,PhysRevX.9.021036,luo2021light,sirica2022photocurrent}, and breaking of time-reversal symmetry by circularly polarized light\cite{Gedik2013Sci,mahmood2016selective,Cavalleri2020NP} to induce valley-selective optical Stark effect\cite{sie2015valley} and Bloch-Siegert effect\cite{sie2017large}. While important progresses have been witnessed for symmorphic symmetry breaking\cite{basov2017towards,SunNRP2021}, nonsymmorphic symmetry breaking still remains largely unexplored. Black phosphorus is a model semiconductor\cite{YeBPACS} where the conduction band (CB) and valence band (VB) can be tuned by a static electric field to generate a band crossing nodal ring\cite{Kim2015sci,Zunger2015NL,Choi2015NL,Quek2015SR,Kim2017prl,Choi20172D} with two gapless nodes protected by a nonsymmorphic symmetry, in particular, the glide-mirror symmetry\cite{Kim2017prl}. Therefore, black phosphorus presents a nice candidate material for investigating the glide-mirror symmetry breaking.

Here, we report the experimental demonstration of a light-induced glide-mirror symmetry breaking in black phosphorous, by utilizing time- and angle-resolved photoemission spectroscopy (TrARPES). Upon near-resonance pumping, the light-field dressed sideband of the conduction/valence band (CB/VB) overlaps with the VB/CB, resulting in a band crossing nodal ring, and a gap is observed along the entire nodal ring. Such observation of fully-gapped nodal ring is in sharp contrast to the application of a static electric field, suggesting breaking of the glide-mirror symmetry. Moreover, the gap co-develops with the light-field in the time domain, suggesting the ultrafast and coherent nature of the light-induced glide-mirror symmetry breaking through Floquet engineering. 

\section{RESULTS AND DISCUSSION}

\begin{figure*}[htb]
	\centering
	\includegraphics[width=16.8cm]{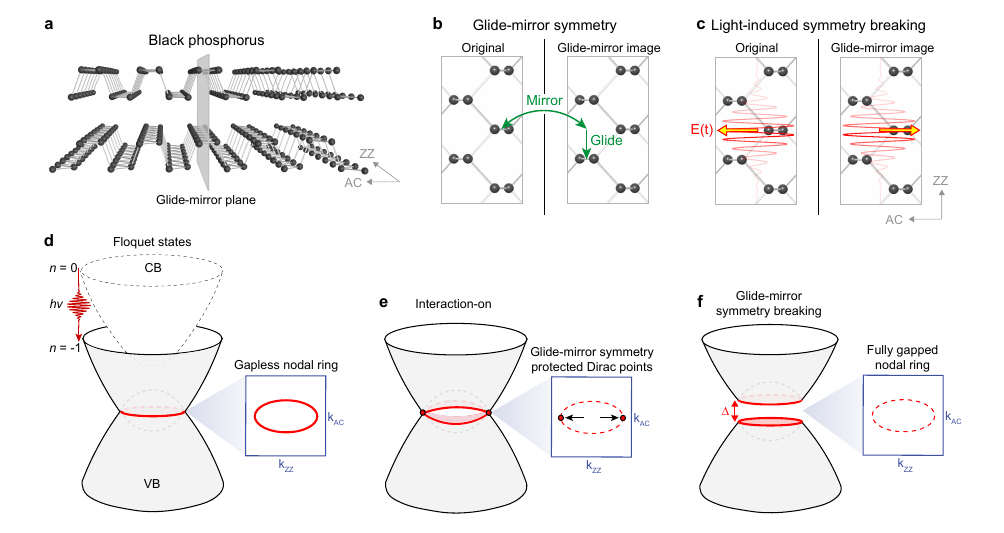}
	\caption{{\bf Glide-mirror symmetry breaking through Floquet engineering in black phosphorus. } (a) The crystal structure of black phosphorus with glide-mirror symmetry. (b) Illustration of glide-mirror symmetry. (c) Illustration of glide-mirror symmetry breaking under the light-field polarized along AC direction. (d-f) Schematic Floquet states of black phosphorus without interaction (d), with interaction (e) and glide-mirror symmetry breaking (f).} 
	\label{fig:schematics} 
\end{figure*}

Black phosphorus crystallizes in a buckled honeycomb lattice with high anisotropy\cite{zhang2014extraordinary,he2015exceptional} as schematically shown in \ref{fig:schematics}a. The sample exhibits a glide-mirror symmetry: it is identical after reflection with respect to the symmetry plane (the grey plane) followed by a translation along the zigzag (ZZ) direction (\ref{fig:schematics}b)\cite{Kim2017prl}. In the Floquet picture,  the light-field-dressed CB Floquet sideband ($n$ = -1) can overlap with the VB, resulting in a band crossing nodal ring\cite{Zhou2023Nature,zhou2023floquet} as schematically shown in \ref{fig:schematics}d. When the light-field preserves the glide-mirror symmetry, the nodal ring might be partially gapped, leaving two glide-mirror symmetry protected Dirac nodes in the glide-mirror plane along the ZZ direction\cite{Kim2017prl,Choi20172D} as schematically shown in \ref{fig:schematics}e, which is in analogous to the equilibrium case where the nodal ring is generated by strain\cite{Neto2014PRL,Li2015PRB,Chen2015PRL,Zou2016PRB,Fang2016PRB,Peeters2016PRB,David20172D,Meng2018prl} or static electric field\cite{Kim2015sci,Zunger2015NL,Choi2015NL,Quek2015SR,Kim2017prl,Choi20172D}. However, when the light-field polarization is perpendicular to the glide-mirror plane (parallel to the armchair (AC) direction), the transient light-field could, in principle, break the glide-mirror symmetry as illustrated in \ref{fig:schematics}c, resulting in gapped Dirac nodes along ZZ direction and a fully gapped nodal ring (\ref{fig:schematics}f). Such a light-induced full gap opening is a key step toward the realization of Floquet topological insulator\cite{Lindner2011natphy}.

\begin{figure*}[htb]
	\centering
	\includegraphics[width=16.8 cm]{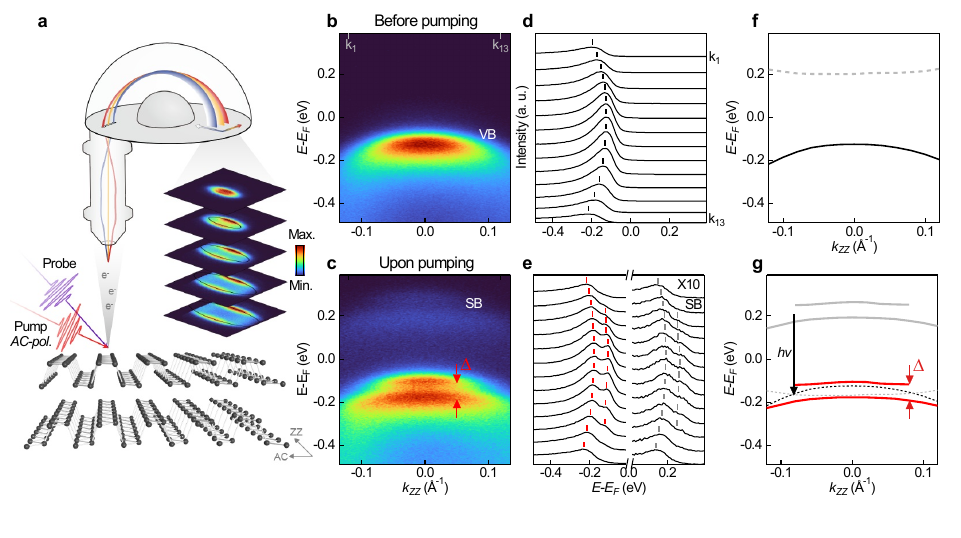}
	\caption{{\bf Light-induced gap opening along the ZZ direction suggesting the glide-mirror symmetry breaking.} (a) Schematics for TrARPES on black phosphorous. (b-e) The dispersion images before and upon pumping (b,c), and the corresponding energy distribution curves (d,e). (f) Extracted dispersion for the VB from the data shown in panel b. The dispersion of the CB (extracted from TrARPES data measured at $\Delta t$ = 1 ps shown in Figure~S1) is added as a dashed curve.   (g) Extracted dispersions from the data shown in panel c. The pump photon energy is 380 meV and pump fluence is 0.7 mJ cm$^{-2}$. The pump polarization is parallel to the AC direction.} 
	\label{fig:gap} 
\end{figure*}

Systematic TrARPES measurements are performed on black phosphorus (\ref{fig:gap}a) to investigate whether the glide-mirror symmetry can be broken through Floquet engineering. A near-resonance (slightly above-gap) pumping light with polarization perpendicular to the glide-mirror plane ($AC$-$pol.$) is applied to generate the band crossing nodal ring. TrARPES measurements are performed along the ZZ direction to reveal the glide-mirror symmetry protected gapless Dirac nodes. Before pumping, the VB shows an overall parabolic dispersion in \ref{fig:gap}b. Upon pumping, it evolves into two well-separated branches as shown in \ref{fig:gap}c, together with another set of bands displaced by a pump photon energy. The dispersions before and upon pumping are extracted from the energy distribution curves (EDCs) analysis in \ref{fig:gap}d,e, which clearly shows a light-induced band gap opening along the ZZ direction in \ref{fig:gap}f,g (see more analysis in Supporting Information and Figure~S1). This experimental observation clearly shows that the Dirac nodes from band crossing are gapped upon pumping.

The wave functions around the CB and VB edges are well-defined superpositions between the atomic orbitals of different sublattices: $\psi_{CB} = \psi_{A} + \psi_{B}$ and $\psi_{VB} = \psi_{A} - \psi_{B}$ (Ref.~\cite{Kim2020natmater}), where  A and B are the two different sublattices in the unit cell of black phosphorus. Upon the glide-mirror symmetry operation (here we use $\mathcal G$ to denote such operation), the A and B sublattices are exchanged (Figure~S2). Applying such operation to the wavefunctions, we obtain $\mathcal{G}(\psi_{CB}) = \psi_{B} + \psi_{A}= \psi_{CB}$ and $\mathcal{G}(\psi_{VB}) = \psi_{B} - \psi_{A} = -\psi_{VB}$. Therefore, the CB and VB edges have opposite eigenvalues for the glide-mirror symmetry operation, so the Dirac nodes in the glide-mirror plane (along the ZZ direction) are protected by the glide-mirror symmetry. Therefore, the observation of gapped Dirac nodes along the ZZ direction indicates the breaking of glide-mirror symmetry.

\begin{figure*}[htb]
	\centering
	\includegraphics[width=16.8 cm]{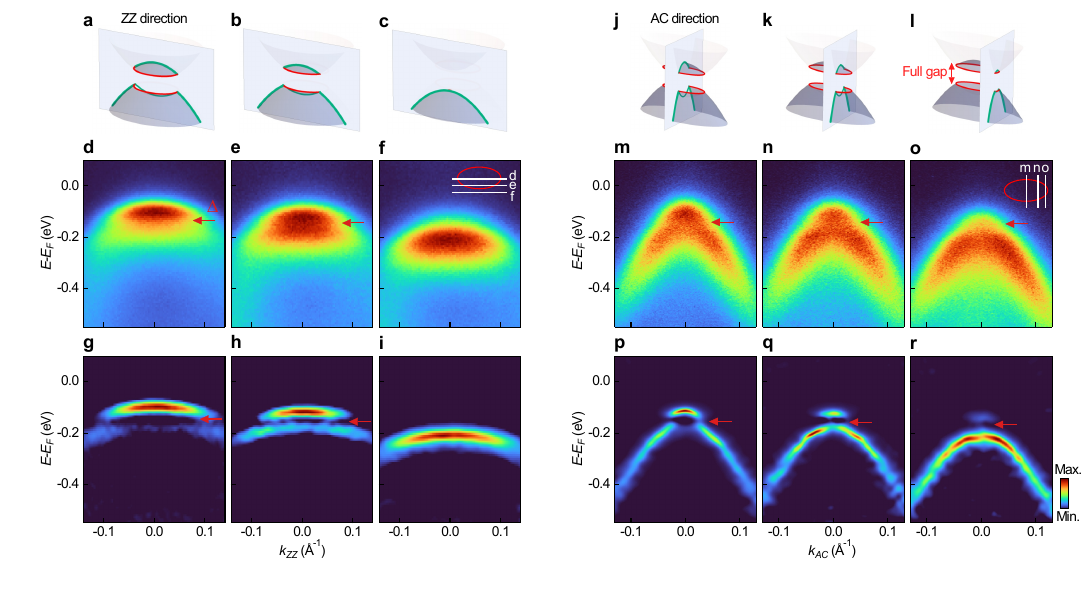}
	\caption{{\bf The observation of fully gapped nodal ring induced by glide-mirror symmetry breaking.} (a-c) Schematic drawing of the measurement slices for data in panels d-f. (d-i) TrARPES dispersion images (d-f) parallel to ZZ directions at k$_{AC}$ = 0, 0.03, 0.06 \AA$^{-1}$ as indicated in a-c and the corresponding second-derivative images (g-i). The red arrows point to the hybridization gap. The pump photon energy is 420 meV and the pump fluence is 0.7 mJ cm$^{-2}$. The pump polarization is along the AC direction. (j-r), Similar data as (a-i) for measurements along AC direction at k$_{ZZ}$ = 0, 0.09, 0.13 \AA$^{-1}$. The pump photon energy is 420 meV, and the pump fluence is 0.9 mJ cm$^{-2}$. The pump polarization is along the AC direction.} 
	\label{fig:FullyGap} 
\end{figure*}

To explore whether the nodal ring is fully gapped, the evolution of the nodal ring is measured in full 2D momentum space. First, a series of dispersion images parallel to the ZZ direction at different $k_{\rm AC}$ (schematically illustrated in \ref{fig:FullyGap}a-c) are measured as shown in \ref{fig:FullyGap}d-f. When the measurement slice crosses the nodal ring (represented by the red ellipses) in \ref{fig:FullyGap}a,b, a clear band gap opening is observed as pointed by the red arrows in \ref{fig:FullyGap}d,e, which can be seen more clearly in the corresponding second-derivative images in \ref{fig:FullyGap}g,h. When the measurement slice derives from the nodal in \ref{fig:FullyGap}c, the gap disappears as expected in \ref{fig:FullyGap}f,i. Moreover, the evolution of dispersion images along the AC direction in \ref{fig:FullyGap}j-r also shows a clear band gap opening as long as the measurement slice crosses the nodal ring. Therefore, the nodal ring is confirmed to be fully gapped in full 2D momentum space, thus confirming the light-induced glide-mirror symmetry breaking again. In contrast, when the pump polarization is rotated to parallel the glide plane, no gap is observed (Figures~S3, S4) due to the reservation of glide-mirror symmetry.

\begin{figure*}[htb]
	\centering
	\includegraphics[]{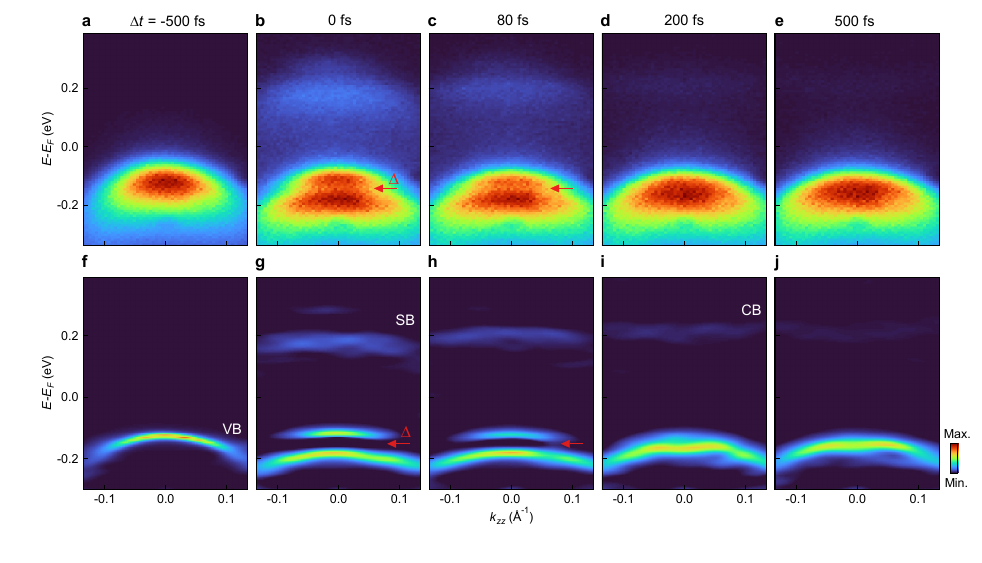}
	\caption{{\bf The ultrafast nature of glide-mirror symmetry breaking.} (a-j) TrARPES dispersion images at different delay times along the ZZ direction (a-e) and the corresponding second-derivative images (f-j). The pump photon energy is 380 meV and the pump fluence is 0.7 mJ cm$^{-2}$. The pump polarization is along the AC direction. The red arrows point to the hybridization gap.} 
	\label{fig:time} 
\end{figure*}

The observed light-induced symmetry breaking enables opportunities for manipulating electronic structures on an ultrafast timescale\cite{ulstrup2016ultrafast,wang2016ultrafast,suzuki2019ultrafast} and potential photoelectronic applications if the gap can be tuned to the Fermi energy, where the high on-off speed plays an important role. To explore how fast the light-induced glide symmetry breaking here, the temporal evolution of the electronic structure is explored. The TrARPES dispersion images at different delay times are shown in \ref{fig:time}a-e, which show that the gap opening only exists around $\Delta t$ = 0 as pointed by the red arrows in \ref{fig:time}b,c and disappears at $\Delta t$ $>$ 80 fs in \ref{fig:time}d,e. This can be seen more clearly in the corresponding second-derivative images in \ref{fig:time}f-j. In the meantime, the photon-dressed sidebands also only appear around $\Delta t$ = 0 as shown in \ref{fig:time}g,h, leaving a long-lived CB at $\Delta t$ $>$ 80 fs (\ref{fig:time}i,j). Therefore, the gap opening co-develops with photon-dressed sidebands, i.e., the light-field,  on a femtosecond timescale, indicating the ultrafast and coherent nature of the observed glide-mirror symmetry breaking through Floquet engineering, which might be applicable for future terahertz or even petahertz electronics\cite{garg2016multi}.

Finally, we would like to discuss some implications of the observed light-induced glide-mirror symmetry breaking. First, since the net electric field could be canceled by averaging over several optical cycles, the glide-mirror symmetry is only broken transiently in the subcycle timescale as schematically illustrated in \ref{fig:schematics}c. Therefore, when the TrARPES measurement is averaged over several optical cycles during the duration of the probe laser, whether any observable effect can be detected remains elusive. Here, the observation of the fully gapped nodal ring induced by such transient glide-mirror symmetry confirms that such transient symmetry breaking can really induce observable effects even averaging over several optical cycles. While similar phenomena have been reported but with different mechanisms, such as optical rectification\cite{bass1962optical} and nonlinear phononics\cite{forst2011nonlinear,mankowsky2014nonlinear,mankowsky2016non}, this work highlights the role of Floquet engineering in revealing and utilizing such transient symmetry breaking. Second, while the full closing and reopening of the band gap is a key route for realizing nontrivial topology\cite{hasan2010colloquium}, it is also key on the road to searching for the Floquet topological insulator\cite{Lindner2011natphy} (the non-equilibrium version of topological insulator). Here, we demonstrate the band crossing nodal ring can be fully gapped experimentally through Floquet engineering, although how to make the light-driven black phosphorus topological is still an open question\cite{PhysRevLett.120.237403}. Third, the symmetry in valley-related materials also plays an important role, for example, time-reversal symmetry for transition-metal dichalcogenide (TMDC) materials\cite{schaibley2016valleytronics} and crystalline symmetry for TiSiCO-family\cite{yu2020valley,cui2023four}. Experimental investigation of symmetry manipulation through Floquet engineering in these materials would be an important direction in the future.  For example, valley-specific Floquet topological phase has been proposed in TMDC materials\cite{sie2015valley}, meanwhile Floquet engineering induced topological phase transition and manipulation of Chern number have been recently proposed in TiSiCO-family\cite{liu2024photoinduced}.

\section{CONCLUSIONS}

In conclusion, a light-induced ultrafast glide-mirror symmetry breaking is demonstrated in black phosphorous through Floquet engineering, leading to the light-induced fully gapped nodal ring.  Such fully-gapped nodal ring formed by overlapping the VB/CB sideband and CB/VB is a key step toward the realization of dynamic Floquet topological insulator\cite{Lindner2011natphy,Lindner2020NRP}. Therefore, our work not only demonstrates Floquet engineering as an effective way to break the nonsymmorphic symmetry, but also moves forward towards steering the topological properties of nonsymmorphic topological semimetals\cite{Fu2015PRB,Wen2015PRB,KanePRL2015,Nodalchain2016,BernevigNat2016} and searching the long-sought Floquet topological phases\cite{Lindner2011natphy,Lindner2020NRP}.

\section{METHODS}

Black phosphorus single crystals were grown by chemical vapour transport method with the same procedure in Ref.~\cite{Zhou2023Nature,zhou2023floquet}. Millimeter-size and high-quality black phosphorus single crystals were obtained. The crystal is cleaved $in~situ$ and measured in an ultrahigh vacuum better than 5$\times 10^{-11}$ Torr and at a temperature of 80 K with liquid-nitrogen cooling. TrARPES measurements were performed at Tsinghua University with the same TrARPES system in Ref.~\cite{Zhou2023Nature,zhou2023floquet}. The probe beam is linearly polarized along the AC direction with a photon energy of 6.2 eV. The pump beam is also linearly polarized along the AC direction, and the pump photon energy and pump fluence are indicated in corresponding figure captions, respectively.

\section{AUTHOR INFORMATION}
\subsection{Author Contributions}
Shuyun Z. conceived the research project. C.B., Shaohua Z., F.W., Haoyuan.Z and T.L. performed the TrARPES measurements and analyzed the data. F.W. and Haoyuan.Z. grew the black phosphorus single crystal. C.B. and Shuyun Z. wrote the manuscript, and all authors commented on the manuscript.
\subsection{Notes}
The authors declare no competing financial interest.

%%%%%%%%%%%%%%%%%%%%%%%%%%%%%%%%%%%%%%%%%%%%%%%%%%%%%%%%%%%%%%%%%%%%%
%% The "Acknowledgement" section can be given in all manuscript
%% classes.  This should be given within the "acknowledgement"
%% environment, which will make the correct section or running title.
%%%%%%%%%%%%%%%%%%%%%%%%%%%%%%%%%%%%%%%%%%%%%%%%%%%%%%%%%%%%%%%%%%%%%
\begin{acknowledgement}

We thank Peizhe Tang and Benshu Fan for useful discussions. This work is supported by the National Natural Science Foundation of China (Grant Nos.~52388201, 12234011 and 92250305), National Key R\&D Program of China (Grant Nos.~2021YFA1400100 and 2020YFA0308800). Changhua Bao acknowledges support from the Project funded by China Science Foundation (Grant No.~BX20230187) and the Shuimu Tsinghua Scholar Program.

\end{acknowledgement}

%%%%%%%%%%%%%%%%%%%%%%%%%%%%%%%%%%%%%%%%%%%%%%%%%%%%%%%%%%%%%%%%%%%%%
%% The same is true for Supporting Information, which should use the
%% suppinfo environment.
%%%%%%%%%%%%%%%%%%%%%%%%%%%%%%%%%%%%%%%%%%%%%%%%%%%%%%%%%%%%%%%%%%%%%
\begin{suppinfo}

Detailed analysis for band overlapping positions, schematics for illustrating sublattice exchange upon the glide-mirror symmetry operation, and experimental data for pump polarization along the ZZ direction.

\end{suppinfo}

%%%%%%%%%%%%%%%%%%%%%%%%%%%%%%%%%%%%%%%%%%%%%%%%%%%%%%%%%%%%%%%%%%%%%
%% The appropriate \bibliography command should be placed here.
%% Notice that the class file automatically sets \bibliographystyle
%% and also names the section correctly.
%%%%%%%%%%%%%%%%%%%%%%%%%%%%%%%%%%%%%%%%%%%%%%%%%%%%%%%%%%%%%%%%%%%%%

\providecommand{\latin}[1]{#1}
\makeatletter
\providecommand{\doi}
  {\begingroup\let\do\@makeother\dospecials
  \catcode`\{=1 \catcode`\}=2 \doi@aux}
\providecommand{\doi@aux}[1]{\endgroup\texttt{#1}}
\makeatother
\providecommand*\mcitethebibliography{\thebibliography}
\csname @ifundefined\endcsname{endmcitethebibliography}
  {\let\endmcitethebibliography\endthebibliography}{}

\end{document}